# Active suppression of temperature oscillation from a pulse-tube cryocooler in a cryogen-free cryostat: Part 1. Simulation modeling from thermal response characteristics

Changzhao Pan[1,2, a)], Bo Gao[1,3,4, a) *], Yaonan Song[1,3, 5], Haiyang Zhang[1,3] Dongxu Han[5], Jiangfeng Hu[1,3,4], Wenjing Liu[1,3,5], Hui Chen[1,3,6], Mark Plimmer[1,2], Fernando Sparasci[1, 2], Ercang Luo[1,3,4], Laurent Pitre[1,2]

[1] *TIPC-LNE Joint Laboratory on Cryogenic Metrology Science and Technology, Chinese Academy of Sciences (CAS), Beijing 100190, China*

[2] *LCM-LNE-Cnam, 61 rue du Landy, 93210 La Plaine-Saint Denis, France*

[3] *Key Laboratory of Cryogenics, Technical Institute of Physics and Chemistry, Chinese Academy of Sciences, Beijing, 100190, China*

[4] *University of Chinese Academy of Sciences, Beijing 100490, China*

[5] *Beijing Institute of Petrochemical Technology, Beijing 102617, China*

[6] *Xi'an Jiaotong University, Xian 710049, China*

**Abstract:**

A cryogen-free cryostat cooled using a 4 K commercial GM or pulse tube cryocooler (PTC) displays temperature oscillations caused by the intrinsic working principle of the regenerative cryocooler. To dampen such oscillations usually requires either a large heat capacity or a large thermal resistance. To understand this phenomenon better and suppress it more effectively, both the step response characteristic and the intrinsic oscillation characteristic of cryostat have been used to obtain the complete transfer functions of a simulation model. The latter is used to test and optimize traditional PID feedback control. The results showed this approach has almost no effect on the temperature oscillation amplitude. Based on this simulation model, a novel active method was proposed and tested numerically. Simulation results predict the method should suppress the amplitude of the original temperature oscillation by a factor of two.

* Author to whom correspondence should be addressed. Email: bgao@mail.ipc.ac.cn;

a) These authors contributed to the work equally and should be regarded as co-first authors

**Nomenclature**

| | |
|---|---|
| *A*, *B* | Temperature amplitudes (K) |
| *C* | Heat Capacity (J·kg$^{-1}$·K$^{-1}$) |
| *G(s)* | Transfer function |
| *L(s)* | Laplace transform |
| *M* | Mass (kg) |
| *Q* | Heat (W) |
| *R* | Thermal Resistance (K·W$^{-1}$) |
| *t* | Time (s) |
| *T* | Temperature (K) |

Greek letters

| | |
|---|---|
| α, θ | Phase angle (rad) |
| ω | Angular frequency (rad·s$^{-1}$) |
| τ | Time constant (s) |

## 1. Introduction

Primary gas thermometry (PGT) methods, such as constant-volume gas thermometry (CVGT) [1], dielectric-constant gas thermometry (DCGT) [2], acoustic gas thermometry (AGT) [3] and single-pressure refractive-index gas thermometry (SPRIGT) [4], measure thermodynamic temperature with the highest accuracy. The related instruments need to measure the pressure or some other parameters of the working gas (dielectric constant, speed of sound *etc.*) at a stable temperature for one to obtain the thermodynamic temperature. For example, SPRIGT has the potential to yield an uncertainty of 0.25 mK in the range 5 K to 25 K via the measurement of a resonant microwave frequency. However, it also requires a very high temperature stability working environment with a standard uncertainty below 0.2 mK. More generally, a cryostat with ultra-high temperature stability is essential for all types of PGT.

Cryostats working at temperatures well below 77 K such as those used in DCGT [2] and AGT [3] often use liquid helium as the refrigerant. However, the need to refill them with helium results in the experiment being discontinuous while the changing liquid gives rises to constantly evolving operating conditions. Moreover, the boiling of liquid helium adds a large thermal noise to the experimental system, especially in the temperature range above 25 K, which makes stable operation of the cryostat difficult [3]. In recent years closed-cycle cryocoolers such as the commercial 4K Gifford-McMahon (GM) cryocooler or the GM-type pulse tube cryocooler (GM-PTC) have been used as an alternative. Cryostats cooled this way, also called cryogen-free cryostats, can be compact and run uninterrupted for several months or even years. They can be used from room temperature down to around 3 K. Naturally, they have attracted increasing interest [5] [6] [7].

At low temperatures (especially below 77 K), however, systems cooled by both the 4K GM cryocooler and the GM-PTC show temperature oscillations caused by the periodic compression and expansion within the cryocooler. At temperatures below 20 K, because of the low heat-capacity of the material within the cryostat, this oscillation becomes even more manifest, thereby hindering temperature regulation.

Thus far, two main approaches have been employed to attenuate this oscillation, namely, the heat-capacity method (HCM) and the thermal-resistance method (TRM). The HCM uses a material of large heat-capacity at low temperature, such as a pressurized helium pot, lead, or a rare-earth alloy (such as ErNi and Er3Ni). Li *et al.* [8], Okidono *et al.* [9] and Webber *et al.* [10] attached a helium filled pot onto the cold end of a GM cryocooler, thereby reducing the peak-to-peak temperature oscillation at 4 K from a few hundred millikelvin to below 50 mK. Fluhr *et al.* and [11], Huang *et al.* [12] used a lead plate to reduce the oscillation amplitude to the millikelvin level. Allweins *et al.* [13] and Shen *et al.* [14] used ErNi alloy with a GM-PTC and attenuated the amplitude to around 10 mK at 4 K. However, while the HCM can suppress the oscillation with almost no increase in lowest working temperature, the additional heat capacity it adds significantly increases the cooling time (typically by an order of magnitude).

The other approach, the TRM, uses a material with low thermal-conductivity at low temperature. Using a fibre-reinforced-plastic (FRP) damper, Nakamura *et al.* [15] reduced the oscillation amplitude at 4.2 K to 0.7 mK while Dong *et al.* [16] using a polytetrafluoroethylene (PTFE) sheet, brought the thermal oscillation amplitude at 20 K down to below 4 mK. Finally, Dubuis *et al.* [17], using lead plates with stainless steel spacers to combine both methods, obtained a temperature stability better than 1 mK. While the TRM does not increase the cooling time, it does cause a temperature jump between the cryocooler and the flange it cools, which leads to a loss of cooling power (by as much as several watts, depending on the thermal resistance).

In addition to the two methods above, in our previous experiment [18] [19] [20], a heat-switch was used to attenuate the temperature modulation.

All the aforementioned techniques, while effective, are passive, their performance depending on the structure of the cryostat. Moreover, their performance cannot be modified in mid-experiment. For this reason, we have sought instead to use active temperature stabilisation. The most commonly used active method is Proportional-Integral (PI) control. To optimise regulation, researchers have developed many theories on tuning PI coefficients [21] [22]. Even so, the study of feedback control with thermal oscillation is rare. Bhatt *et al.* [23] used the transfer function of a cryocooler to help tune the PI control coefficients of their system. To model the thermal behavior of their cryocooler, Sosso *et al.* [24] used linear system theory, which can be effective in the design of temperature controllers. Nevertheless, both these groups' research was only focused on the cold end of the cryocooler itself, and not the cryostat it was cooling. Complete mastery of active control requires that one develop a model for the entire system. Only then can one expect to obtain optimal suppression of temperature oscillation in a cryogen-free cryostat. In the present work, such a model was developed and implemented. It allowed the thermal characteristics of the system to be deduced by experiment. Moreover, it highlighted the inefficacy of traditional PI control and led us to invent a different method for removing thermal oscillations more effectively (by a factor of two).

The remainder of the paper is structured as follows. First, the cryostat and cooler are briefly described. The simulation model is presented, followed by the results obtained when it was implemented with the experiment allowing the determination of all the thermal transfer functions. Finally, the novel active suppression method is described.

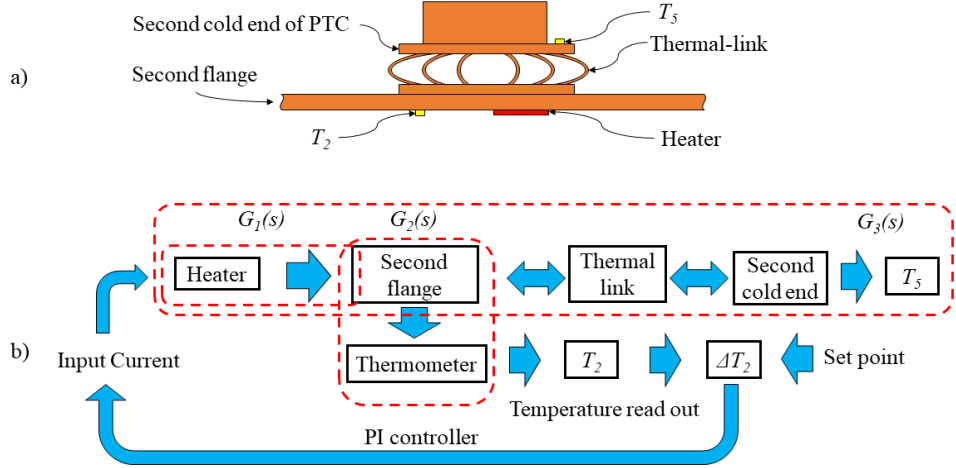

Figure 1. a). Schematic diagram of the experimental system b). Schematic diagram of the PI temperature control feedback loop. Define PTC: pulse-tube cryocooler. The labels $T_2$ and $T_5$ refer to two of the many thermometers placed throughout the system [18], [19].

## 2. Cryogenic apparatus

The cryostat studied in this paper was that used in our system for SPRIGT, a detailed description of which can be found in references [18] and [19]. It was cooled by a commercial Sumitomo GM-type pulse tube cryocooler (PTC), which can yield a lowest temperature of 2.8 K, can supply 1 W cooling power at 4.2 K. Figure 1 shows the structural diagram of the main components studied in this work. A flexible thermal-link was used to connect the second cold end of PTC and the second flange. The heat capacity of the flange and the contact resistances of the thermal link provide passive damping of the oscillation from the PTC. Despite this, the temperature of the second flange still exhibits an oscillation with a peak-to-peak amplitude of several millikelvin. Moreover, these thermal oscillations can diffuse into the pressure tube and pressure vessel and thereby degrade the stability of *pressure* regulation too[1].

---

[1] The microwave resonator for SPRIGT is housed within a pressure vessel, the latter connected to a pressure control system at room temperature via a long tube. Wherever this tube passes through a flange, a thermal-link was added to reduce its temperature. So any temperature oscillation on the flange would be transferred to the pressure tube, thereby causing the pressure to oscillate and degrading the stability of pressure regulation.

To regulate the temperature of the second flange, a heater and two calibrated thermometers (Lakeshore Cernox model CX1050-CU-HT-1.4L[2]) were used.

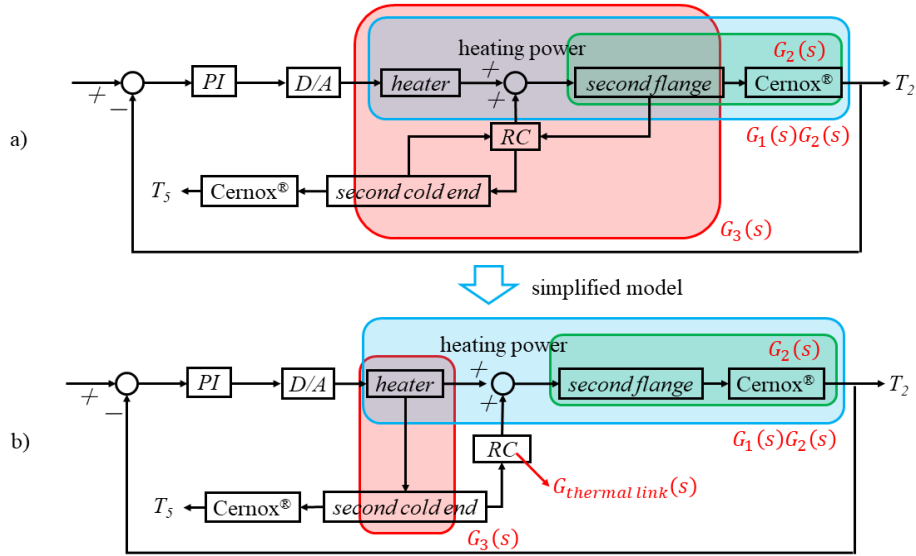

Figure 2 a). Actual temperature control system b). Simplified representation used to model it.

## 3. Simulation model

A schematic of the real temperature control system is shown in figure 2(a). To regulate the temperature of the cryostat better and reduce this oscillation, a simplified temperature control system block diagram was developed (Figure 2(b)). Two factors influence the temperature of the flange: the cooling power from the second cold end of the PTC and the heat supplied by the heater. However, the heater also warms the second cold end of the PTC. To simplify the simulation model, the heat transfer paths are reduced to four modules (figure 2b). The first module concerns the heat transfer from the heater to the flange, modelled by the function $G_1(s)$. The second module involves the heat transfer from the flange to the thermometer, modelled by the function $G_2(s)$. The third, modelled by the function $G_3(s)$, quantifies the heat transfer from the heater to the second cold end of the PTC and then to another thermometer. Finally, the effect of the thermal link is modelled by the transfer function $G_{thermal\ link}(s)$.

---

[2] To simplify the installation of the thermometer, the CU type package was used in the experiment. But the relatively low thermal response time of the CX1050-CU sensor partially attenuated the amplitude of the temperature oscillation. From a relative standpoint, this had no influence on the outcome of the research. However, we suggest the reader use preferably a thermometer with a faster response to measure the dynamic temperature, such as the models CX1050-SD, BC or similar.

These four transfer functions are enough to describe the main thermal response characteristics. To obtain expressions for them, a step response experiment was performed on the flange and the second cold end of the PTC. As shown in figure 3, a heat step (from 0 to 8.8 mW) was added by the heater to the open loop system. The transfer functions $G_1(s)$ and $G_3(s)$ were then obtained by using the ratio of the Laplace transform of the output function to the Laplace transform of the excitation step heating function. Indeed, according to references [22] and [24], the best fit of the temperature response to a power step input can be accurately described by a sum of exponential decays:

$$T(t) = A_0 + A_1\left(1 - e^{-t/\tau_1}\right) + A_2\left(1 - e^{-t/\tau_2}\right) + \cdots \tag{1}$$

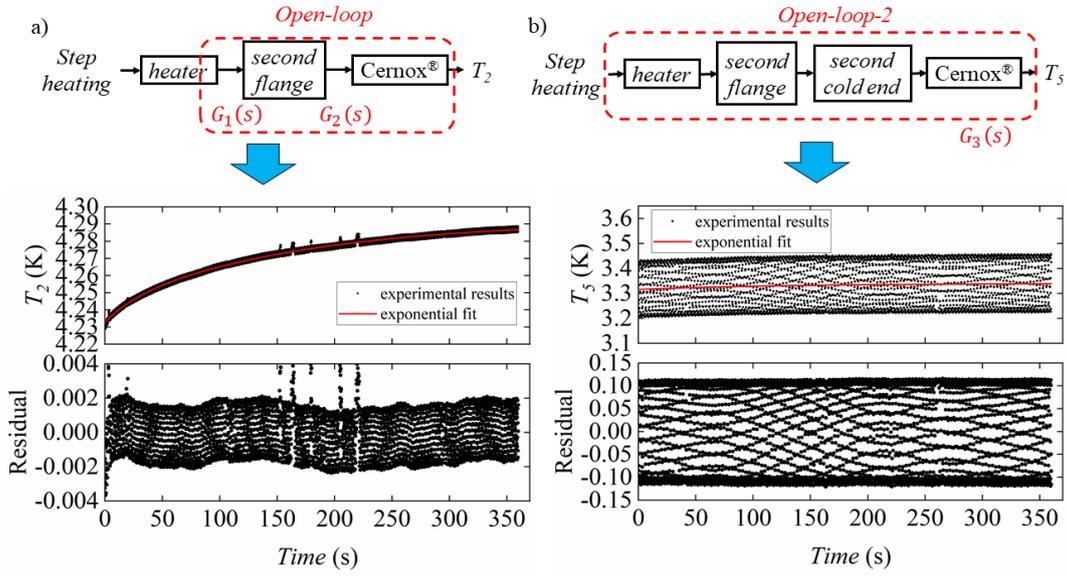

Figure 3. Step responses of a) the flange and b) the PTC second cold end temperatures, $T_2$ and $T_5$ respectively.

Table 1. The fitted functions of the stepwise heating and intrinsic sinusoidal response.

|  |  | Fitted function |
| --- | --- | --- |
| Step response | $T_2$ (K) | $4.23 + 0.01\left(1 - e^{-t/26.48}\right) + 0.05\left(1 - e^{-t/165.41}\right)$ |
|  | $T_5$ (K) | $3.34 - 0.024\left(1 - e^{-t/134.96}\right)$ |
| Sinusoidal response | $T_2$ (K) | $4.26 + 0.0017 \times \sin(10.84t + 1.89)$ |
|  | $T_5$ (K) | $3.33 + 0.12 \times \sin(10.76t + 3.02)$ |

To simplify the model, a first-order response function was used to describe heat transfer from the heater to the second-stage cold end of the PTC. A second-order response function was used to describe that from the flange to thermometer, because it contains two parts $G_1(s)$ and $G_2(s)$ and has two time-constants. Figures 3 a) and b) show the response curves of the flange and the second-stage cold end of the PTC. The fitted response functions can be found in table 1. From its Laplace transform, we obtain the following expressions for the transfer functions:

$$G_1(s)G_2(s) = \frac{340.91s+6.82}{(26.48s+1)(165.41s+1)} \tag{2}$$

$$G_3(s) = \frac{2.77}{135s+1} \tag{3}$$

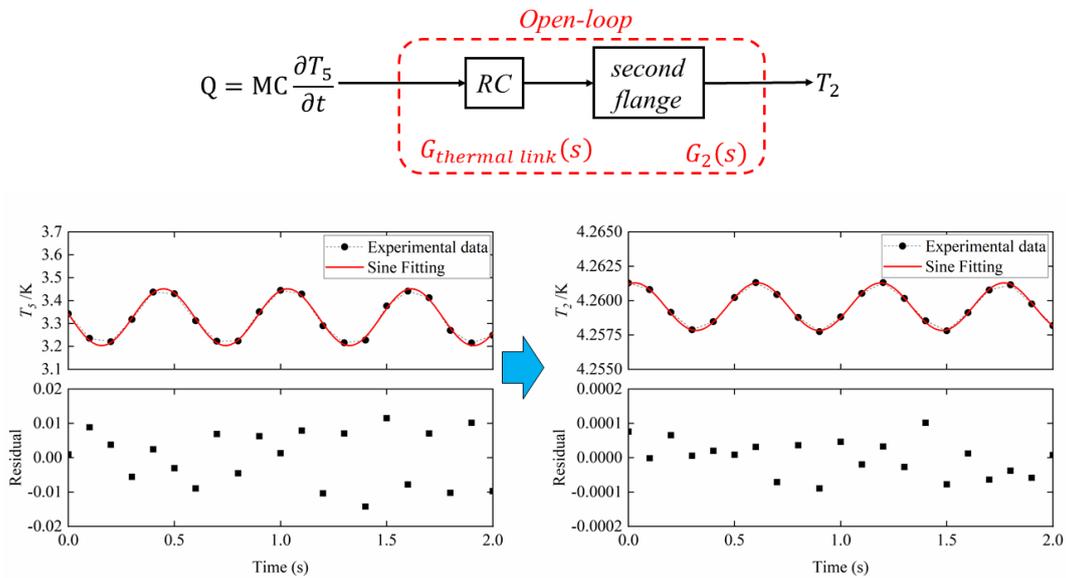

Figure 4 The sinusoidal response from the PTC to the thermometer on the second flange.

In addition to the step response method, the sinusoidal response method can also be used to obtain information on the transfer functions. In the present system, there always exists a temperature oscillation propagaing from the second cold end of PTC to flange[3]. This means there also exists a heat oscillation (around a mean value of zero) travelling between the second-stage cold end of the PTC and the flange. Thus, by analyzing this intrinsic sinusoidal response, one can obtain an expression for transfer function. As shown in figure 4, this heat oscillation can be expressed by the derivative of $T_5$, propagating through the thermal link and the second flange to the thermometer $T_2$. By applying the Laplace transform to this open loop response, one can obtain the complete transfer function from the second cold end of the PTC to the thermometer on the second flange:

$$G_{thermal-link}(s) \cdot G_2(s) = \frac{L(T_2 - A_0)}{L(Q)} = \frac{1}{MC} \frac{\frac{A_1 \sin\theta}{-\omega^2 B_1 \sin\alpha}s + \frac{A_1 \cos\theta}{-\omega B_1 \sin\alpha}}{\frac{\cos\alpha}{-\omega \sin\alpha}s + 1} \qquad (4)$$

---

[3] For the GM type cryocooler, the pressure wave inside of cryocooler is not actually a sine wave, but rather more like a square wave (containing the fundamental and odd harmonics), so the temperature oscillation of the working gas should be non-sinusoidal. However, because of the heat capacity of the cold end, the higher-order harmonics are greatly filtered, resulting in the measured temperature being essentially sinusoidal after all.

Here, the transfer function of the thermal link $G_{thermal-link}(s)$ can be expressed in terms of its time constant $RC$:

$$G_{thermal-link}(s) = \frac{F_{thermal-link}(s)}{RCs+1} \qquad (5)$$

while the transfer function $G_2(s)$ can be written as:

$$G_2(s) = \frac{F_2(s)}{\tau_2 s + 1} \qquad (6)$$

In our previous work, the thermal resistance of the thermal-link was calculated accurately [19]. At this temperature, $RC \approx 100$ ms, so the term $RCs$ can be neglected to begin with in the actual calculation. Then, from equation of (4) (5) (6), we obtain:

$$G_{thermal-link}(s) \cdot G_2(s) = \frac{1}{MC} \frac{\frac{A_1 \sin\theta}{-\omega^2 B_1 \sin\alpha}s + \frac{A_1 \cos\theta}{-\omega B_1 \sin\alpha}}{\frac{\cos\alpha}{-\omega \sin\alpha}s + 1} \approx \frac{F_{thermal-link}(s) \cdot F_2(s)}{\tau_2 s + 1} \qquad (7)$$

On the other hand, there are two time-delays in the product of equation (2), which correspond the time constants of $G_1(s)$ and $G_2(s)$, respectively $\tau_1$ and $\tau_2$. Because $G_2(s)$ is the transfer function describing heat flow from the flange to the thermometer, it should have the smaller time constant. Thus, we take $\tau_2 = 26.48$ s. To solve equation (7), one also needs to know the phase relationship between $\alpha$ and $\theta$ and which can be obtained by measuring $T_2$ and $T_5$ simultaneously. In the experiment, a trigger device was used to control two Keithley 2002 digital multimeters (DMM) to measure both temperatures at the same time. The phase difference $\alpha - \theta$ was found to be 1.134 rad, as shown in table 1.

When the following values of the other parameters are inserted, $M = 1.6$ kg, $A_1 = 0.00173$ K, $B_1 = 0.12361$ K, $\omega = 10.83$ rad s$^{-1}$, equation (7) simplifies to:

$$G_{thermal-link}(s) \cdot G_2(s) = \frac{-0.294s + 1.4732}{(0.1s+1)(26.48s+1)} \qquad (8)$$

Comparing equations (2) and (8), the simplest solutions are:

$$G_1(s) = \frac{340.91s+6.82}{165.41s+1} \tag{9}$$

$$G_2(s) = \frac{1}{26.48s+1} \tag{10}$$

$$G_{thermal-link}(s) = \frac{-0.294s+1.4732}{0.1s+1} \tag{11}$$

## 4. Discussion

In this section, we describe the simulation model developed using MATLAB Simulink software and how simulation results were verified by experiment. Thereafter, we present a novel active suppression method to attenuate the thermal modulation intrinsic to cryogen-free cryostats.

### 4.1 Simulation results and verification

Based on above transfer functions, a simulation model for temperature regulation was developed using MATLAB Simulink software. As shown in figure 5, the heat on the flange can be split into two parts. The first is the heat from the heater, calculated using the transfer function $G_1(s)$. The second is the heat due to the temperature oscillation of the PTC, which can be calculated using the transfer function $G_{thermal-link}(s)$. Both heat sources warm the flange and so influence the thermometer temperature $T_5$ via the transfer function $G_2(s)$. In the simulation, the set point was controlled by a step signal, and the temperature evolution observed on a computer monitor.

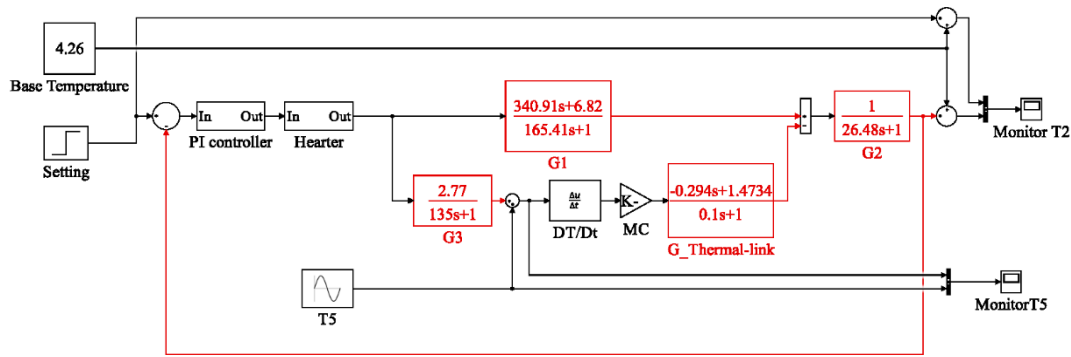

Figure 5 Simulation model written in MATLAB Simulink software

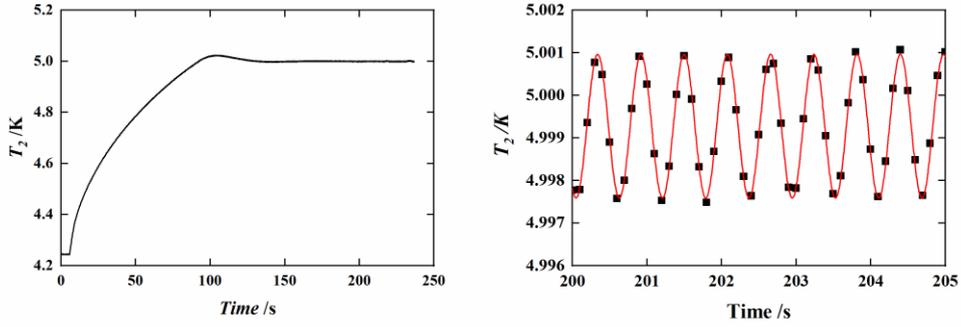

Figure 6. Experimental results for PI regulation of the second flange temperature T2. a) Temperature variation for the first 250 s. After 200 s, the mean value varies by less than 1 mK b). Oscillations around the set point

Figure 6 shows experimental results for temperature regulation at 5 K. In the experiment, the PI method was used to regulate the temperature on the flange, with $Kp = 80$, $I = 4$ and a sampling time of 80 ms. To avoid damage to the heater, the input current to the heater was limited to 100 mA. The results show it took about 150 s for the temperature $T_5$ to increase from 4.23 K to 5.00 K. In the steady state, $T_2$ still shows an oscillation with an amplitude around 1.5 mK.

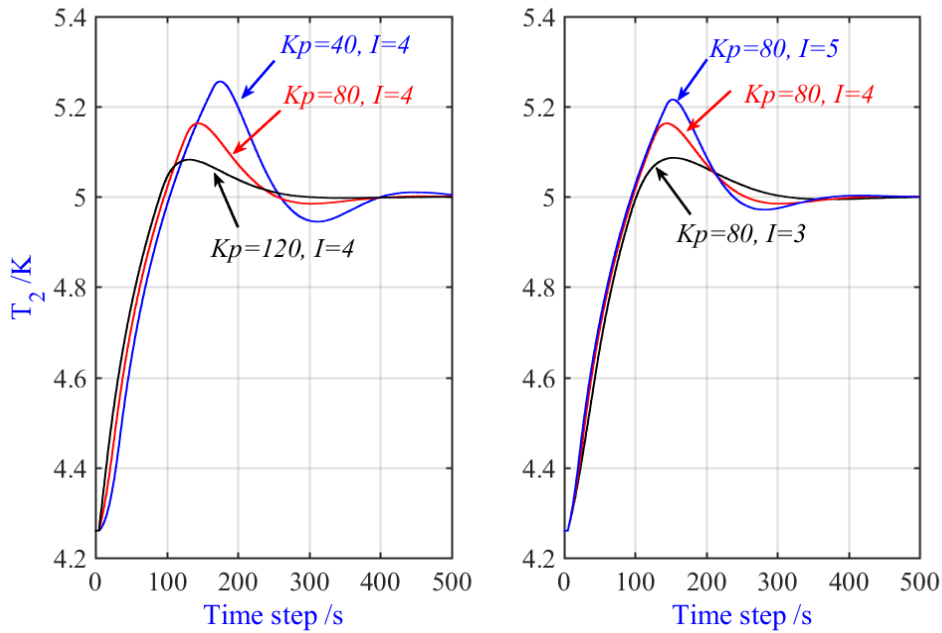

Figure 7. Simulation results with different PI parameters (proportional gain $Kp$ and integral gain $I$) showing temperature evolution towards equilibrium.

Figures 7 and 8 show the results obtained using the present simulation model. One observes that the overshoot value and stable time changed with different PI parameters. Because the maximum current was set at 100 mA in the simulation, a larger proportional gain *Kp* made the temperature reach its stable state more quickly and with a smaller overshoot value. By contrast, a larger integral coefficient *I* caused a larger overshoot value but had little effect on the stablisation time. On the other hand, the PI parameters have almost no effect on the temperature oscillation amplitude (the amplitude of temperature oscillation changed by only 2% at most), which means the oscillation cannot be actively suppressed using traditional PI feedback. If in the simulation, temperatures appear to take a little longer to reach steady state than in the experiment, the final temperature oscillation amplitudes match experimental results well. The difference in stabilisation time might be caused by the changing of thermal diffusivity with temperature. Because the transfer function method is a 1D simulation, it is assumed that the whole system has a uniform temperature. However, in the real 3D experiment, the flange temperature shows some spatial variation. When the temperature changes from one value to another, even if the temperature of thermometer (located near the heater) has changed to that of the set point, it probably takes longer for the temperature to vary at a more distant location. Consequently, the experimental temperatures reach stability more quickly.

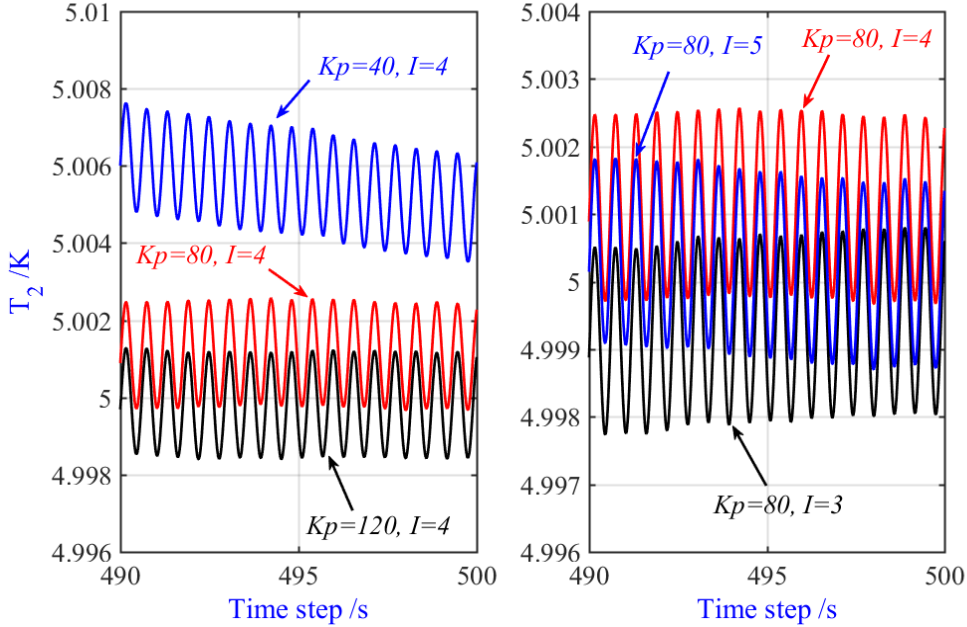

Figure 8. Simulation results with different PI parameters: temperature oscillations around the set point.

**4.2 Active suppression method**

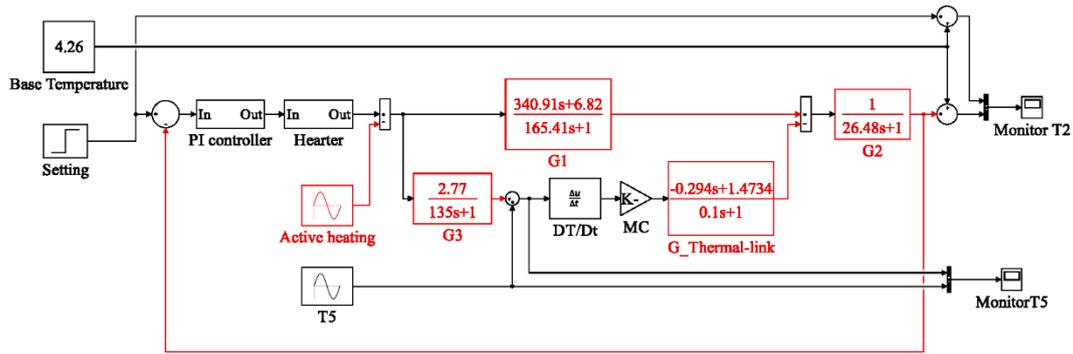

Figure 9 The program used to model active suppression temperature oscillation method.

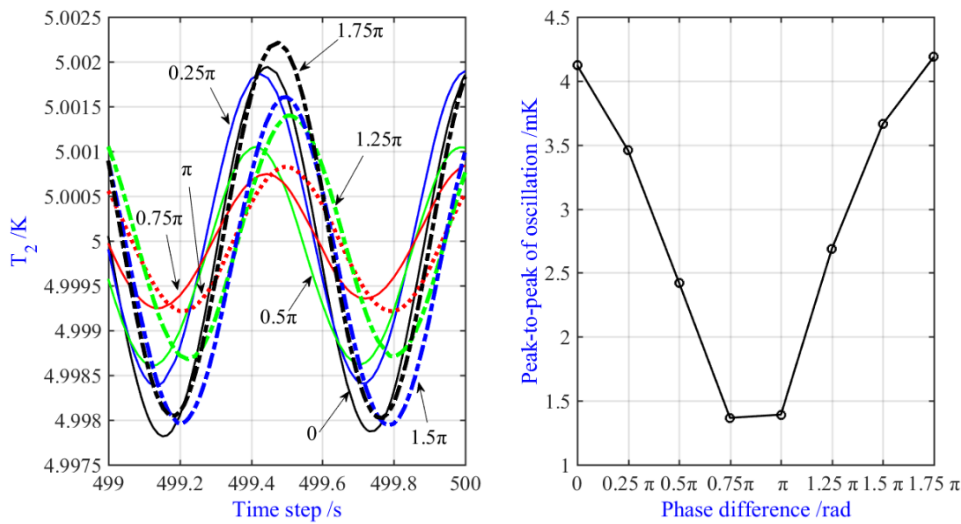

Figure 10 Simulation results for active suppression for eight different phase angles from 0 to $7\pi/4$. The oscillation amplitude is minimized for a phase of between $3\pi/4$ and $\pi$.

Because traditional PI regulation cannot actively suppress the temperature modulation induced by the PTC, we needed to find another way to attenuate or even remove it. The approach used was to add an extra, *active* heat oscillation source to the flange to counterbalance the unwanted thermal oscillation. To test this idea theoretically, a sinusoidal heat source was added onto the previous PI simulation model (figure 9). Because by definition a heater cannot cool, the amplitude of this active sinusoidal heat must be lower than the heat used in the previous PI regulation (in this case around 110 mW).

Theoretically, the addition of two sine-waves can either increase or decrease the total amplitude. Thus, there exists an optimal phase relationship between this active sinusoidal heat and the oscillation heating from PTC which makes the oscillation on the flange lowest. For this reason, several phase angles from 0 to 2π were tested (amplitude set to 100 mW). Figure 10 shows the simulation results. One sees as expected that the original temperature oscillation can be either increased or decreased depending on the phase angle. For a phase angle between 3π/4 and π, the amplitude of the original oscillation is reduced to 1.4 mK. In other words, the amplitude of the original temperature oscillation on the flange (2.75 mK) was suppressed almost two-fold.

## 5. Conclusion

To understand PI regulation with temperature oscillation in a cryogen-free cryostat at low temperature (5 K) cooled by a pulse-tube cooler, a simulation model was developed based on the transfer-function method by using both the step response characteristic and the intrinsic oscillation characteristic of cryostat. The results showed that the thermal response characteristics of the cryostat can be well described by four transfer functions. They also showed that the PI parameters mainly affect the overshoot value and stablisation time during the regulation but have a negligible effect on the thermal oscillation amplitude. In other words, the temperature oscillation cannot be suppressed using PI feedback alone. Based on this observation, a new active suppression method was proposed and tested numerically. According to calculations, it should reduce the original modulation amplitude by a factor of two. The successful experimental implementation of the method will be described in a companion paper submitted to this journal [25].

**Acknowledgments**

This work was supported by the National Key R&D Program of China (Grant No. 2016YFE0204200), the National Natural Science Foundation of China (Grant No. 51627809), the International Partnership Program of the Chinese Academy of Sciences (Grant No. 1A1111KYSB20160017) and the European Metrology Research Programme (EMRP) project：Real-K (No. 18SIB02). Changzhao Pan was supported by funding provided by an H2020 Marie Skłodowska Curie Individual Fellowship-2018 (834024).